\documentclass[fleqn,10pt]{wlscirep}
\usepackage{graphics,soul,hyperref}

\title{Clustered marginalization of minorities during social transitions induced by co-evolution of behaviour and network structure}

\author[1,2,*,+]{Carl-Friedrich Schleussner}
\author[2,3,+]{Jonathan F. Donges}
\author[4,5,+]{Denis A. Engemann}
\author[2,6,7]{Anders Levermann}
\affil[1]{Climate Analytics, Berlin, Germany}
\affil[2]{Potsdam Institute for Climate Impact Research, Potsdam, Germany}
\affil[3]{Stockholm Resilience Centre, Stockholm University,  Stockholm, Sweden}
\affil[4]{Cognitive Neuroimaging Unit, CEA DSV/I2BM, INSERM, Universit\'e Paris-Sud, Universit\'e Paris-Saclay, NeuroSpin center, Gif/Yvette, France}
\affil[5]{Neuropsychology \& Neuroimaging Team, INSERM UMRS 975, ICM, Paris, France}
\affil[6]{Lamont-Doherty Earth Observatory, Columbia University, New York, USA}
\affil[7]{Institute of Physics and Astronomy, University of Potsdam, Potsdam, Germany}

\affil[*]{schleussner@pik-potsdam.de}
\affil[+]{equal contributions}


\begin{abstract}
Large-scale transitions in societies are associated with both individual behavioural change and restructuring of the social network. These two factors have often been considered independently, yet recent advances in social network research challenge this view. Here we show that common features of societal marginalization and clustering  emerge naturally during transitions in a co-evolutionary adaptive network model. This is achieved by explicitly considering the interplay between individual interaction and a dynamic network structure in behavioural selection.
We exemplify this mechanism by simulating how smoking behaviour and the network structure get reconfigured by changing social norms. Our results are consistent with empirical findings: The prevalence of smoking was reduced, remaining smokers were preferentially connected among each other and formed increasingly marginalised clusters. We propose that self-amplifying feedbacks between individual behaviour and dynamic restructuring of the network are main drivers of the transition. This generative mechanism for co-evolution of individual behaviour and social network structure may apply to a wide range of examples beyond smoking.
\end{abstract}

\begin{document}

\flushbottom
\maketitle

Behaviour is shaped by interactions between the individual and its environment~\cite{skinner1981selection}.
As a result of evolutionary pressures emanating from intensified group lifestyles, humans acquired a diverse and specialised social behavioural repertoire~\cite{Dunbar1993,herrmann2007humans}. The human cognitive capacity for enduring collaborative social interaction has been extensively investigated in various disciplines related to the field of cognitive sciences (for an overview cf. ~\cite{tomasello2005understanding,fehr2007social,engemann2012games}). Theories about how behaviours are shaped by social and individual factors have a longstanding tradition in psychology and social sciences~\cite{festinger1962theory,janis1977decision}. Examples of quantitative models include dynamic models of segregation in urban neighbourhoods~\cite{schelling1971dynamic}, models of cultural dissemination~\cite{axelrod1997dissemination} and a wealth of literature aiming at the inclusion of social decision making into economic theory \cite{Akerlof1997,Steinbacher2014}. An overview on mathematical approaches to social dynamics is provided in ref.~\cite{Castellano2009}. 
Importantly, social relations can be represented as graphs, which renders them accessible to network theoretical analysis~\cite{Newman2010}. It has been shown that the overall structure of connections between individuals in social networks and face-to-face interactions between individuals systematically affect a wide range of social and individual characteristics, such as happiness, divorce rates, smoking and obesity \cite{Christakis2007c,fowler2008dynamic, Christakis2008,mcdermott2013breaking,christakis_fowler12}. We refer to this effect as behaviour selection emphasizing an evolutionary process rather than mere individual decision-making.

In this context, networks statistics enable more targeted characterisation of social dynamics. For example, social distance modulates similarity and behavioural synchrony between individuals. It is commonly measured using the shortest path length between two individuals in a social network and has been shown to preferentially shape individual behaviour up to a distance of three social ties \cite{christakis_fowler12}. Likewise, the contents of social interactions between individuals depend on their relationship, i.e., perceived friendship status, which suggests nonlinear interdependencies between network dynamics and social interaction~\cite{Szell2010}.
Opinion formation and imitation have been advocated as candidate mechanisms of behaviour induction \cite{Girvan2002,asch1955opinions}. This puts emphasis on cognitive processes and biases for selective spread of behaviours in social networks \cite{Crandall2008,colman2003}.

These findings motivate process-based techniques for modelling dynamical social networks highlighting time-varying aspects of social connections~\cite{Snijders2010}. One such approach is represented by adaptive networks that model the temporal co-evolution of network structure and dynamic node states~\cite{gross2008adaptive,gross2009adaptive,sayama2013modeling}. The most commonly modelled social processes include imitation, collaboration dynamics and social tie formation as a function of antipathy and sympathy (sometimes referred to as homophily and heterophily, respectively) between agents. Adaptive network models have then been used to investigate complex phenomena in social networks such as phase transitions and tipping points in opinion formation on single-~\cite{Holme2006} and multi-layer~\cite{Diakonava2015} networks, epidemic spreading~\cite{gross2006epidemic}, swarm behaviour~\cite{huepe2011adaptive}, friendship structure in social media networks~\cite{li2013coevolving}, sustainable use of renewable resources~\cite{Wiedermann2015} and coalition formation~\cite{Auer2015}. Opinion formation has been intensively investigated using the adaptive voter model~\cite{Holme2006}, its generalisations and related models~\cite{gross2008adaptive,gross2009adaptive,sayama2013modeling} with a focus on consensus formation~\cite{Nardini2008}, opinion diversity~\cite{demirel2011cyclic} and network fragmentation~\cite{bohme2011analytical}.

In adaptive network models, the selection of update rules critically determines the co-evolutionary dynamics of node states and network structure. When considering real-world social systems, social connections are very unlikely to be established randomly and in disregard of the underlying network structure, but rather are the outcome of agents' interactions in a complex network~\cite{bozon1989finding,Kossinets2006a,Henry2011}. At the same time, interaction between agents along social ties is a key process to induce individual behavioural change \cite{hegselmann2002}. Therefore, choosing update rules such that they explicitly take into account peculiarities of micro-scale social interactions seems promising for obtaining more realistic models. Such an interaction-resolved adaptive network approach would then also allow to study the co-evolution of micro-scale social influence and large-scale network structures. 

More empirical findings have become available that explicitly describe social network structures. In the work presented here, we will focus on a study on smoking habits by Christakis and Fowler \cite{Christakis2008}.
Based on a detailed long-term survey, they analysed smoking habits of 12,067 inhabitants of a small town in the US between 1971 and 2003 while concomitantly tracking their social relationship structure, i.e., mutual assessment of friendship status. Their analysis revealed that over that time period, the prevalence of smoking declined from about 50\,\% to about 10\,\%. At the same time, the structure of social connections changed almost selectively for the remaining smokers. Their average eigenvector centrality, a measure of how much a node is in the ``centre'' of its social network, significantly declined. At the same time, the probability of an individual being a smoker conditional on the prevalence of smoking in its neighbourhood (referred to as \emph{conditional probability} below, see Methods) increased up to the level of third degree contacts (contacts of contacts of contacts).
In other words, individuals who did not adapt to the decreasing societal support for smoking preferentially interacted with similar individuals, forming subgroups or clusters of increasingly marginalised smokers.

\begin{figure}[!htbp]
\centering
\includegraphics[width=8 cm]{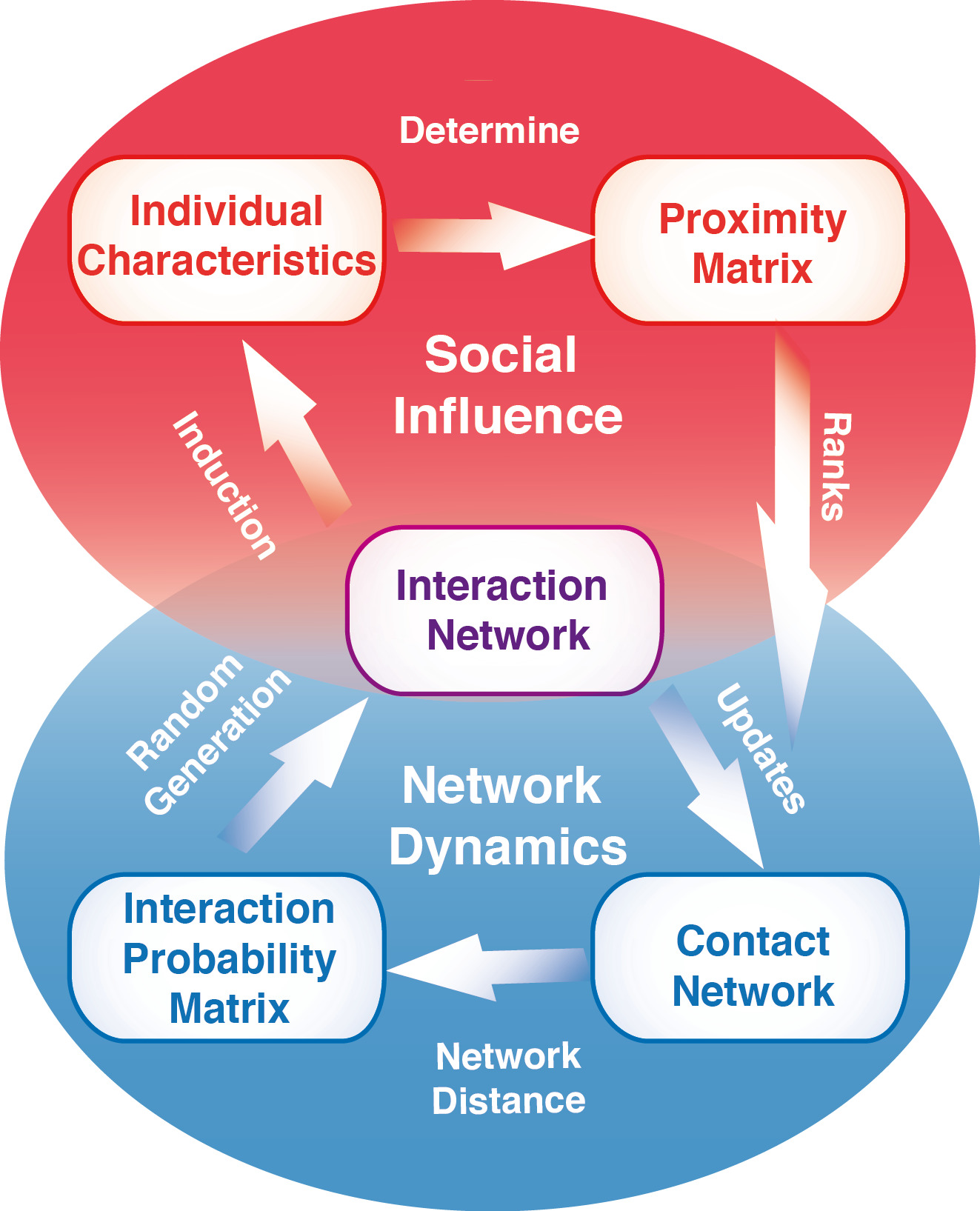}
\caption{\textbf{Adaptive network model of behaviour selection.}
For a group of individuals, the proposed model predicts selection of behaviour as a function of two factors: local interaction between individuals and the global structure of their social connections.
The \emph{proximity matrix} describes how similar a given pair of
individuals is based on their \emph{individual characteristics} such as smoking behaviour.
Assuming restricted resources, agents maintain a limited number of social contacts.
In the proposed model, individuals only keep their most proximate contacts.
Based on the \emph{proximity matrix}, it can be determined which individuals are current neighbours in the
\emph{contact network}. The distance between nodes in this network is then used to compute the \emph{interaction probability matrix} for stochastically generating current interactions between a given pair of individuals. Importantly, this \emph{interaction network} exerts feedback on the individual behaviour and thus closes the co-evolutionary loop: The probability of changing the smoking behaviour is modelled as a function of the individual smoking disposition and the dominance of smoking behaviour in the local neighbourhood of the interaction network.
Note that only individuals who have actually interacted can establish a tie in the contact network in the next time step.
}
\label{fig:scheme}
\end{figure}

\section*{An adaptive network model of behaviour selection}

In the following, we will introduce an interaction-resolved adaptive network approach that we evaluate in terms of its capacity to reproduce characteristics of empirically studied time-varying social networks. We first outline the general modelling framework that contains core conceptual ideas of our adaptive network model of behaviour selection presented in Figure \ref{fig:scheme}. In a next step, we describe specific modifications and additions made to model social dynamics of changing smoking behaviour to reproduce findings from the empirical reference case \cite{Christakis2008}.

Complex systems, such as the human brain, social networks, or the backbone structure of the internet, typically implement functional hierarchies~\cite{van1992information,roberts2015heavy,Watts1998a,dehaene2003neuronal}. Although dynamics with multiple temporal hierarchies also apply to the emergence of complex macroscopic structure in social systems in which agents repeatedly interact over time~\cite{flack2011challenges,flack2012multiple,dedeo2013collective}, hierarchical social network dynamics have rarely been explicitly modelled~\cite{gross2008adaptive,gross2009adaptive,luhmann2015memory,dedeo2010inductive}. Here we considered functional hierarchies as coupling between an interaction network with fast updates and contact network with slow updates that together shape individual characteristics as their states change over time with preferential formation of social ties. A schematic overview of the model and its components is depicted in Fig.~\ref{fig:scheme} and a detailed formal description of our model is given in the methods description below.

The \textit{contact network}'s structure is based on an overall similarity between individual's characteristics such as preferences, socio-economic status or genetic factors (cf. ~\cite{domingue2014genetic,werner1979similarity,eiser1991adolescent,kobus2003peers}) that generate a social proximity between individuals. Here, contacts are understood as the number of other agents an individual may regularly interact with (a counterexample for this are entries in a Facebook contact list that only require a single interaction to be established). The total number of such contacts that can be maintained by a human individual is constrained by temporal and cognitive capacities~\cite{Dunbar1993}. General cognitive capacity and the number of contacts are both subject to individual differences~\cite{bickart2012intrinsic,deary2010neuroscience}.
We therefore restricted the maximum degree of social contact that an agent in the contact network is capable or willing to maintain by introducing an individual \textit{degree preference} parameter that is normally distributed. An agent cannot maintain more contacts than prescribed by its degree preference, which implies that establishing new contacts (by a new edge in the contact network) may require disbanding old ones (deleting the edge in the contact network).

 The \textit{interaction network} provides the basis for establishing new contacts while at the same time also inducing change in the individual characteristics.
It is generated stochastically at each time step based on the contact network. Reflecting empirical findings \cite{christakis_fowler12}, the probability of interactions between two individuals in a given time step (represented by an edge in the interaction network) decreases with the shortest path distance between them in the contact network. The minimal interaction probability is a constant positive value, thereby allowing for unlikely, incidental meetings (``by chance'') between distant or disconnected individuals. 

In our model, tie formation in the contact network is constrained by the social proximity that may change as a result of the interaction \cite{Onnela2010}. To update the contact network, the social proximities to neighbours in the interaction network are compared with proximity values to contact network neighbours. Only the top ranking contacts are maintained, both in the contact and the interaction network up to the agent specific \textit{degree preference}. Importantly, to establish a contact between two agents, each of them has to be included in the other's set of preferred contacts. By this requirement of reciprocity, previous contacts can be replaced actively, but also lost passively, reminiscent of forgetting. Such a process can be illustrated by an agent moving from city \textit{A} to another city \textit{B} in which she establishes new contacts and at the same time gradually forgets about her previous social network in A. In turn, also her previous contacts in \textit{A} loose contact with her.

At the same time, the social influence dynamics play out on the interaction network. Individual characteristics such as behaviours are subject to peer-influence by direct neighbours in the interaction network as will be described below. The model design also allows to account for individual dispositions as node-dependent constraints on behaviour exogenous to the model, i.e., weights on choice options, that do not depend on the interaction network. Such dispositions may be understood as culturally transmitted norms, values, knowledge and slowly changing collective contexts (e.g. health campaigns or climate change). Intuitively, by altering these weights according to a simulation protocol, one can emulate changes in global societally relevant factors. 
The hierarchical coupling between components in our model supports decomposition into partial models (see Table~\ref{table} and Methods). This allows us to differentiate the relative importance of model components and their associated social processes for behaviour selection in response to changing global trends.

\begin{table}[!tbp]
\centering
\caption{Partial and alternative models of behaviour selection studied in this work.}
\label{table}
\begin{tabular}{llll} \toprule
 &  & \multicolumn{2}{l}{behaviour update by social interaction} \\ \cline{3-4}
 &  & yes & no \\
feedback into contact network & yes & \emph{coupled}, \emph{mean-field} & \emph{network} \\
 & no & \emph{interaction} & -- \\ \bottomrule
\end{tabular}
\end{table}

\section*{Modelling social dynamics of changing smoking behaviour}

In the following, we apply the proposed adaptive network model of behaviour selection to the specific case of network-dependent changes in smoking behaviour to investigate empirically observed social transitions.

\begin{figure}[!htbp]
\centering
\includegraphics[width=16 cm]{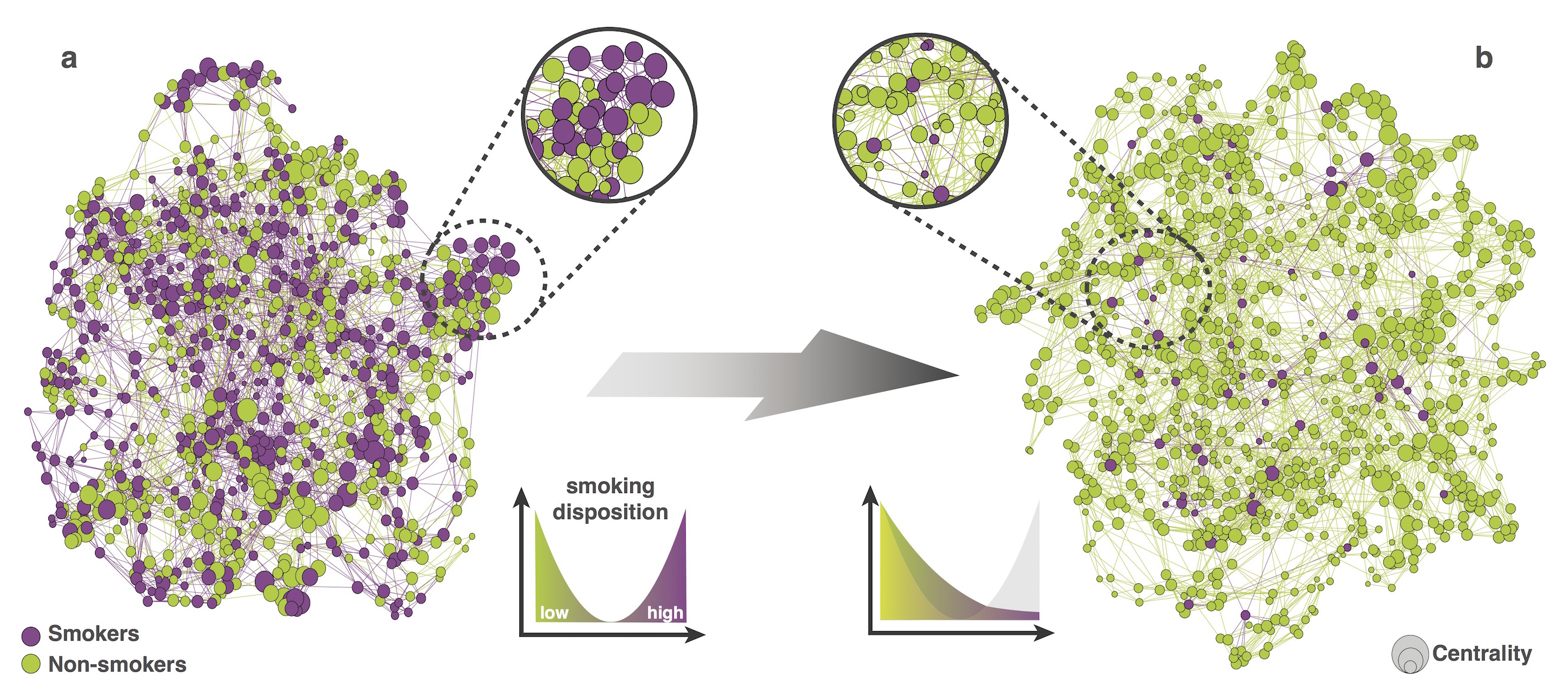}
\caption{\textbf{Smoking behaviour and centrality before and after the social transition.}
Panel (a) and (b) illustrate the initial and the final state of the contact network as simulated by the proposed adaptive network model of behaviour selection.
Circles represent individual nodes. Their colour and size represent smoking behaviour and the individual's centrality, respectively. At the initial state of the simulation (panel (a)), smoking behaviour is homogeneously distributed across the network with random centrality values and the number of smokers equals the number of non-smokers resulting from the initial distribution of smoking dispositions. As the normative support for smoking gradually declined, behaviour and centrality changed over repeated interactions within the network. In the final state of the simulation (panel (b)), the number of smokers has considerably declined. As a consequence of the adaptive network dynamics, the centrality of smokers is selectively reduced. In comparison, non-smokers are characterised by a wide distribution of centrality.}
\label{fig:example}
\end{figure}

In particular, we introduce an update mechanism for the agent's smoking behaviour as the individual characteristic of interest. We conceptualise \textit{smoking behaviour} as a binary variable (either smoking or non-smoking) endogenously in the model. The individual's smoking behaviour can be altered over time by an Ising-type model of social influence \cite{Castellano2009}. At each time step, we determine the probability of an agent to alter its smoking behaviour as a function of balanced peer-influence of smoking and non-smoking behaviour of its neighbours in the interaction network (see Methods). In addition to the peer-influence, we introduce an \textit{individual smoking disposition} that reflects individual preferences in the probability to switch smoking behaviour. As this individual smoking disposition is exogenous to the model, its distribution can be altered externally and the dynamic response of the model can be investigated. 

Importantly, it is only the endogenous binary smoking behaviour that dynamically affects agents' social proximity in our model. As in actual social networks, however, the social proximity also reflects many dimensions of which most remain latent during an interaction. Our \textit{proximity matrix} thus includes two components: the time-invariant \emph{background proximity} that largely determines the position of agents in the social network, e.g. reflecting long-term social ties such as family relationships, and a time-dependent component that depends on the co-occurrence of smoking behaviour for pairs of individuals. As a consequence, adopting a new behaviour will modify an agent's entries in the proximity matrix and may thus lead to changes in the contact network.

We then emulated dynamics of societal changes that historically lead to reduced prevalence of smoking (e.g. health campaigns and changes in public opinion~\cite{Christakis2008}) by gradually modifying the exogenous distribution of smoking disposition.
Over 1000 model time steps, we gradually converted a bimodal distribution, representing a balanced share of smokers and non-smokers in the network, into a quasi unimodal distribution favouring non-smoking attitudes as depicted in Fig.~\ref{fig:example}. We subsequently performed simulations over an ensemble of 1000 model runs using different seeds to initialise the pseudo-randomisation of the time-invariant background proximity matrix and the smoking disposition (see Methods). The model dynamics of interest were robust with respect to the specific choice of the distributions and the speed of the change in external forcing.

\begin{figure}[tb]
\centering
\includegraphics[width=12 cm]{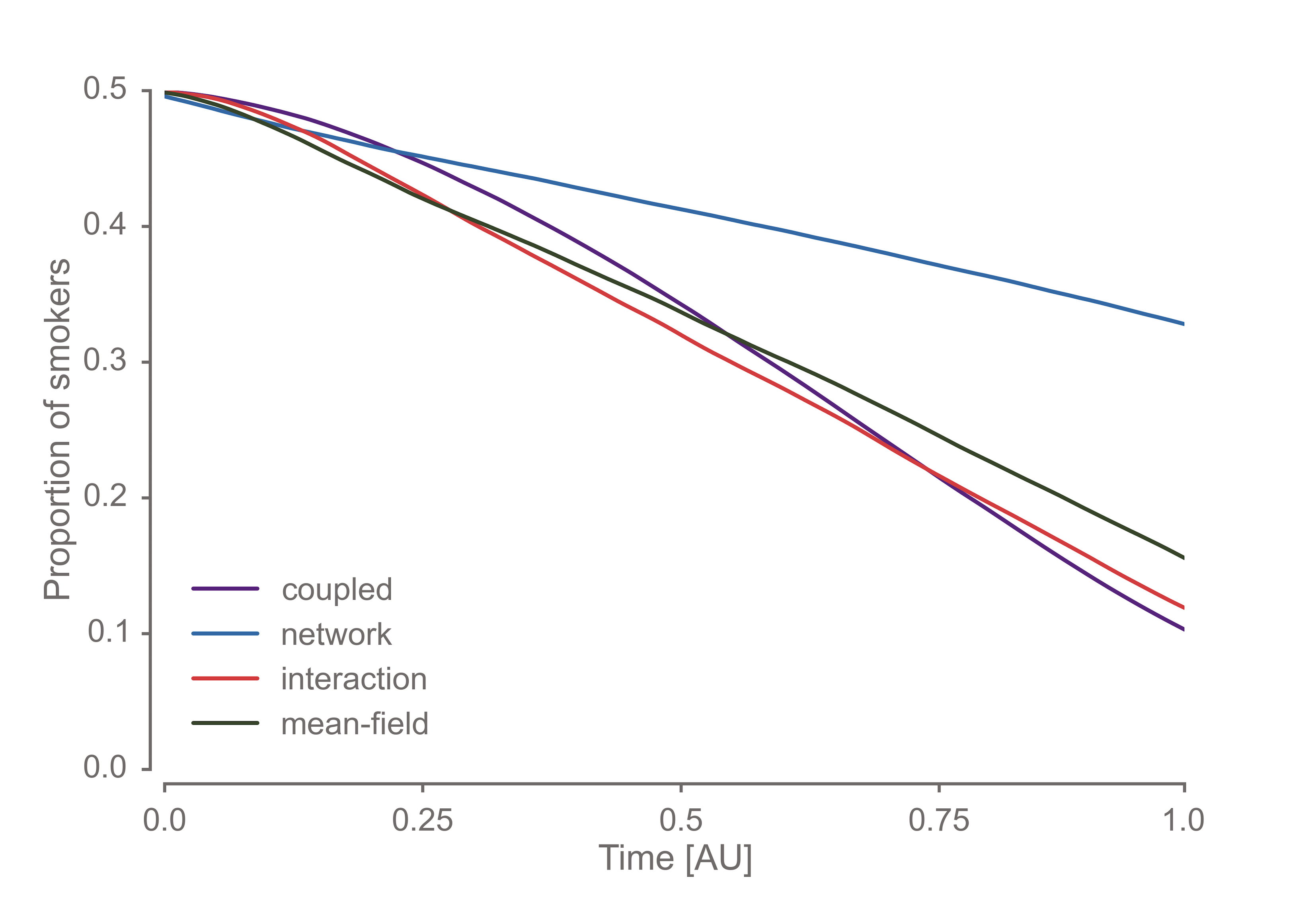}
\caption{\textbf{Gradual transition from a smoker to a non-smoker society is reflected in the prevalence of smoking.}
We considered four distinct models of behaviour selection. Over the course of the simulation, the distribution of the smoking disposition was gradually transformed from a bimodal to a quasi unimodal distribution and the smoking behaviour was computed at each time step.
The \emph{coupled} model assumes that behaviour is shaped by a local interaction based on a time-varying contact network.
In the \emph{network} model, no local interactions are considered. Here, the behaviour is only determined by the individual disposition while the contact network changes over time. In the \emph{interaction} model, only local interactions shape the behaviour on a static contact network. Similarly to the \emph{coupled} model, social influence and network dynamics are considered in the \emph{mean-field} model, but smoking behaviour is shaped by non-local influences only. In all four (partial) models, the proportion of smokers changes in the course of the normative transition (time is represented in arbitrary units, AU). Notably, the absolute change is higher in the models that assume social interaction. Solid lines show mean values across 1000 runs with different pseudo-random initializations. The variation across runs was negligible.}
\label{fig:n_smokers}
\end{figure}

\section*{Results}

To evaluate our co-evolutionary model of behaviour selection, we gradually modified the exogenous smoking disposition and studied the response of several metrics of our social adaptive network. These metrics were motivated by previous empirical studies that documented co-evolution between behaviour and social network structure (cf. ~\cite{christakis_fowler12,Christakis2008}) and include the prevalence of smokers in the network, the eigenvector centrality of each individual and the probability that an individual smokes given that her contacts smoke (conditional probability of smoking).

Over time, the fraction of smokers in the network reduces from about 50\,\%  to 10\,\% (Fig.~\ref{fig:n_smokers}), which is consistent with the empirical findings reported in~\cite{Christakis2008}. In a second step, we compared different generative mechanisms by repeating the analysis for the remaining three partial models (see Tab.~\ref{table} and \emph{coupled} model in Fig.~\ref{fig:n_smokers}). We found that models considering social influence dynamics (\textit{interaction}, \textit{mean-field} and \textit{coupled}) reduced the smoking prevalence twice as much as the network model (Fig.~\ref{fig:n_smokers}), in which agent's behaviour is determined solely by the exogenous smoking disposition.

\begin{figure}[tbp]
\centering
\includegraphics[width=12 cm]{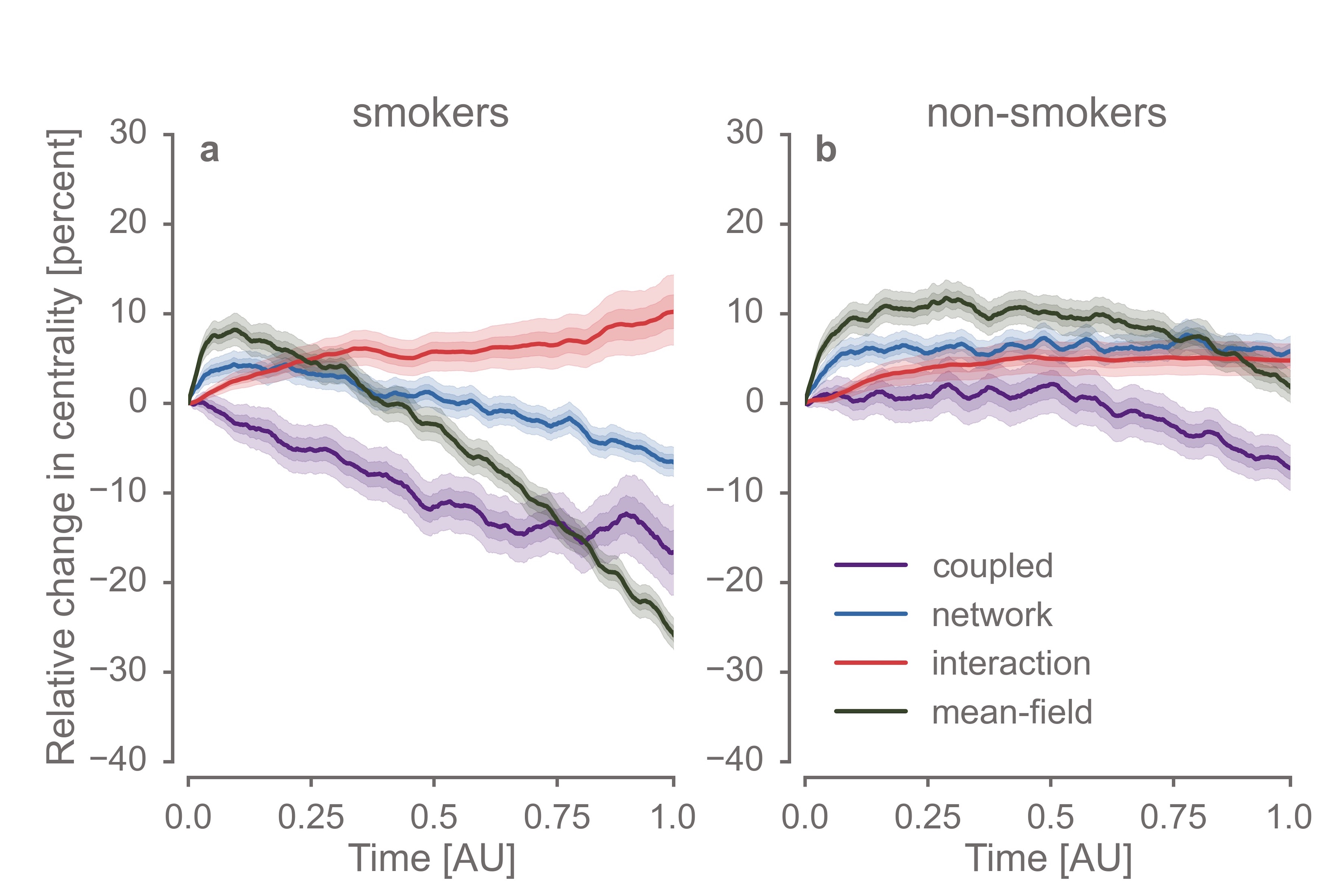}
\caption{\textbf{During the normative transition, local interactions between individuals and the evolution of the network structure rendered smokers less influential in the network.}
The eigenvector centrality (EVC) was computed at each time step for 1000 model runs with different random seeds. All models were initialised to the equilibrium run of the coupled model and values were normalised to the initial state of the model parameters. Panel (a) and (b) depict changes in EVC according to the four (partial) models for smokers and non-smokers, respectively.
Solid lines show mean values and areas indicate bootstrapped 95\,\% and 99\,\% confidence intervals. It is noteworthy that only in models reflecting network effects,  EVC was substantially reduced. These effects were strongest in models that in addition considered social interactions between individuals.}
\label{fig:centrality}
\end{figure}

\begin{figure}[tbp]
\centering
\includegraphics[width=12 cm]{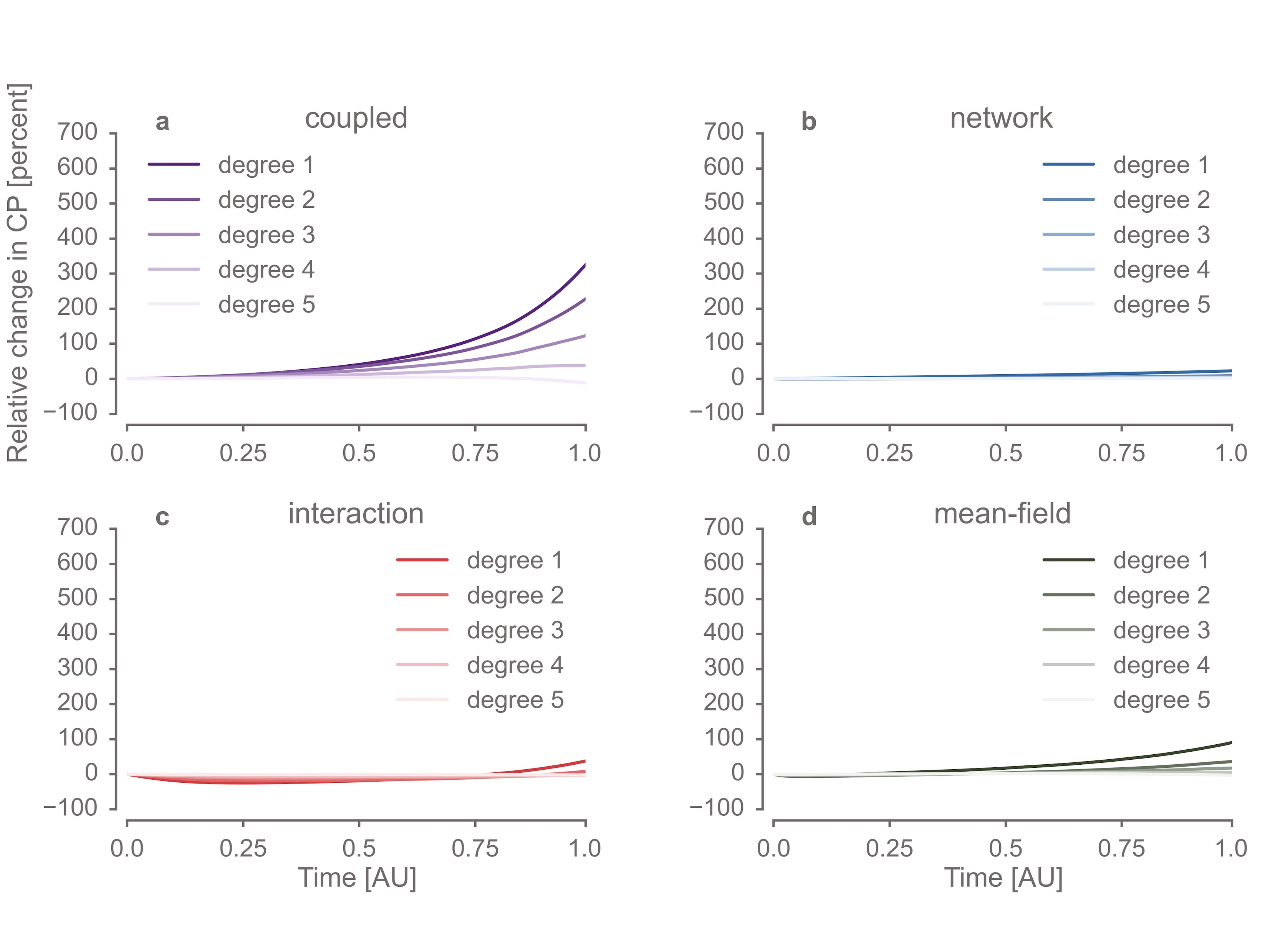}
\caption{\textbf{During the normative transition, joint effects of individual interaction and the evolution of the network structure led remaining smokers to cluster in marginalized groups.}
The probability of an individual being a smoker conditional on the prevalence of smoking in its neighbourhood at a given social distance or \emph{degree} of separation
(conditional probability, CP) was computed for 1000 model runs with different pseudo-random seeds at each time step and normalised to the percentage of change relative to the initial value. Panels (a--d) show the CP at social distance 1--5 for the \emph{network}, the \emph{interaction}, the \emph{coupled} and the \emph{mean field} models, respectively. CP increased with simulated time for lower social distances (1--3) across all models, whereas decreases of CP were observed for larger social distances in some models. Notably, this pattern was between four to eight times more pronounced in the \emph{coupled} model compared to all other models.
Solid lines depict mean values and areas represent bootstrapped 99\,\% and 95\,\% confidence intervals, barely visible as a result of the high signal to noise ratio for this metric.}
\label{fig:conditional_proba}
\end{figure}

Only models considering the evolution of the contact network reduced the eigenvector centrality of remaining smokers below baseline (Fig~\ref{fig:centrality}a). These effects were most pronounced for models combining social influence and network dynamics (\textit{mean-field} and \textit{coupled}). These models suggest a preferential reduction of centrality for smokers, reminiscent of the empirical results reported in~\cite{Christakis2008} (Fig~\ref{fig:centrality}b). Here the \textit{mean-field} model exhibits somewhat more drastic effects with less temporal variability as compared to the \textit{coupled} model and even initially increases the eigenvector centrality of nodes, however not selectively for smokers. This is consistent with the deactivation of local influence that might give rise to ''clusters of resistance''.
When considering the conditional probability of smoking up to fifth degree contacts (Fig.~\ref{fig:conditional_proba}), we found changes between four to eight times higher in the~\emph{coupled} model as compared to all other models. This suggests that the feedback between specific \emph{local} dynamics of social influence greatly amplifies such social clustering behaviour. Taken together, the results from our fully~\emph{coupled} co-evolutionary model support key findings from the empirical reference study~\cite{Christakis2008}. At the same time, these results suggest that the residual pattern of clustered smokers of reduced centrality reflect a synergy between local interaction and network dynamics. 

\section*{Discussion}

We proposed a co-evolutionary model of behaviour selection in adaptive social networks and evaluated it through computational models targeting historical changes of smoking behaviour in social networks.
Our computational models emulated gradual changes of network-wide smoking norms. We observed a reduced prevalence of smoking, a decreased eigenvector centrality and an increased conditional probability of smoking. Notably, the patterns of smoking behaviour and network characteristics computed by our model closely resemble empirical findings from a large-scale and long-term social network study investigating smoking behaviour in a North American small town~\cite{Christakis2008}.
Results of a partial model analysis suggest that selective modelling of either network dynamics, social influence or non-local social induction yields less match with empirical findings, and underscore the empirical relevance of behaviour-network co-evolution. Only the fully coupled co-evolutionary model was capable of explaining non-trivial structural change in complex social systems. 

In particular, we would like to highlight the relevance of local generative mechanisms for social interaction as indicated by the deviating results for a~\textit{mean-field} forcing. The apparent imminent relevance of locality in interactions underscores the need for meaningful, social network based update mechanisms to study complex social phenomena.


It is important to highlight that our models did neither involve any data-fitting nor predictive analyses. Instead we provided simulations with outputs according to their parameters and components. Thus the reported evidence is qualitative in nature and emphasises one distinct generative model through comparisons to empirical data and prior knowledge. The variance across ensembles assumes different values metric-wise, reflecting their algebraic properties as well as the effect of network size. Hence, our analyses do not imply statistical inference. The specific parameter choices in our model were adapted from the empirical study \cite{Christakis2008} and motivated from social sciences, evolutionary biology and neurosciences findings~\cite{Dunbar1993,buzsaki2014log,linkenkaer2001long,domingue2014genetic,christakis_fowler12}. 


Furthermore, the models we evaluated in our simulations clearly suffer from conceptual limitations. The cognitive make-up of our simulated individual constitutes a bold simplification, particularly the assumption of an exogenously prescribed behavioural disposition. Human behaviour is clearly not binary but continuous and more complex model assumptions can therefore be easily motivated. Decision making is governed by multiple interacting factors, involving individual cognitive-emotional dispositions, but also by collaboration dynamics integrating social and cultural factors. Social support for a certain behaviour is often ambiguous, reflecting conflicting values, and social interactions can be asymmetric and unequally weighted.
In this context, our model of behavioural change should be regarded as a prototype. We do not assess the validity of a specific mechanism of behavioural change or opinion formation. Instead we emphasize the structural importance of co-evolutionary processes that coalesce social cognition with network dynamics. But we hope that our simulation method stimulates future validation of specific social cognition theories against the background of evolving social networks.
%
Nevertheless, our model generalises to other empirically documented examples of behaviour-network co-evolution including the spread of happiness, the spread of obesity but also the conditioning of food choices \cite{pachuki11}, as well as large data sets available from monitored social dynamics in massive multiplayer online games~\cite{Szell2010}. Assessing behavioural changes in social networks thereby complements spatial analysis \cite{gallos2012collective}, as social ties and physical distance tend to be substantially correlated \cite{Christakis2008} albeit a spatial and a network approach highlight fundamentally different qualities of the environment.

In particular, the example of food choices illustrates the potential outreach of our model for diverse interdisciplinary research questions. For example, the environmental foot-print of meat-centred diets is considerably higher than that of a vegetarian diet~\cite{stehfest2009climate}. Against the historical background of strong positive correlations between meat consumption and economic prosperity~\cite{Popp2010451} and given the rise of the global middle-class, diet habits represent a key challenge affecting several planetary boundaries~\cite{Rockstrom2009,steffen2015planetary}. At the same time, modifying nutritional behaviour has been targeted by health-related disciplines such as clinical psychology and behavioural medicine in preventive and therapeutic contexts. Capitalizing on contingencies between individual behaviour and environment, therapeutic efforts might therefore benefit from models that specifically detail the relationship between microscopic and macroscopic social dynamics. Furthermore, co-evolutionary dynamics of behaviour and networks have been found to promote cooperation in public good games~\cite{Fowler2010,fehl_2011}, thereby illustrating their transformative potential \cite{benn2014organizational}. In the neuroscientific context it would be worthwhile to explore adaptive networks of spontaneous brain activity. Such models might help to overcome the limitations of ubiquitous "flat models" which do not resolve functional and structural connectivity hierarchically. 

At a theoretical level, our study promotes a synergistic, co-evolutionary and interdisciplinary approach to social dynamics which gains explanatory momentum by integrating interpersonal cognitive processes with network dynamics as explanatory factors.

\section*{Methods}


\subsection*{Detailed model description}

In the following, we provide a detailed mathematical model description, including definitions of the model variables and parameters, their initialisation, the algorithm for computing their temporal dynamics, the general modelling protocol and partial models. The model code is publically available and can be assessed and reviewed here: \url{https://github.com/pik-copan/pycopanbehave}. The implementation is based on the Python complex network software package \texttt{pyunicorn}~\cite{donges2015unified} that is available at \url{https://github.com/pik-copan/pyunicorn}.

\paragraph{Model entities} We model individuals as \emph{agents} or nodes $i \in V$, where $V$ denotes the population or set of $N=|V|$ agents in the social system considered. The system is assumed to be closed and, hence, $V$ does not depend on time, i.e. agents cannot enter or leave the population. The following entities define the system on the individual and population levels:

\paragraph{Agent properties} Each agent $i$ carries a vector of scalar \emph{agent properties}. Agent properties are parameters prescribed externally, they are fixed at initialisation and can be changed over time only by forcing external to the model (see below). The current model setup implements two agent properties:

(i)~\emph{Degree preference} $q_i \leq N-1$ is a discrete quantity serving as an upper bound of an agent's degree in the contact network $k_i^C$ (i.e. its number of neighbours in the contact network). This reflects the varying and limited capability of individuals to establish, manage and maintain sustained social relationships~\cite{Dunbar1993}. $q_i$ is drawn from a discretised Gaussian distribution with mean $\mu$ and standard deviation $\sigma$ at initialisation of the model. Specifically, we choose $\mu=10$ and $\sigma=3$ in the smoking experiment.

(ii)~\emph{Smoking disposition} $\gamma_i(t) \in \left[0,1\right]$ is a continuous variable measuring an agent's individual and network-independent preference for smoking. Agents with small smoking disposition $\gamma_i(t)$ have a low probability to start smoking if they are non-smokers, while those with large $\gamma_i(t)$ have a low probability to stop smoking if they are smokers. $\gamma_i(0)$ is drawn at initialisation from a bimodal, parabolic probability density distribution that is optionally modified by stochastic external forcing towards an quasi unimodal distribution over time (see below).

\paragraph{Agent characteristics} Each agent $i$ also carries a vector of scalar \emph{agent characteristics}. The latter are dynamic variables that are by definition subject to social influence (induction), i.e. they are internal variables of the model obeying the social influence loop (Fig.~\ref{fig:scheme}). In this work, the following single agent characteristic is implemented: \emph{smoking behaviour} $s_i(t) \in \left\{0,1\right\}$ is a binary variable describing an agent's actual smoking behaviour. $s_i(t)=0$ means that an agent does not smoke at time $t$, $s_i(t)=1$ implies that an agent does smoke at time $t$.

\paragraph{Contact network} The \emph{contact network} resembles the social relationships between agents. By contacts we refer explicitly to people that interact on a regular basis and are able to follow each other's affairs. We describe it as an undirected and simple time-dependent graph $G^C(t)$. It can be represented by its adjacency matrix $\mathbf{A}^C(t)$. The neighbourhood of agent $i$ in the contact network is denoted by $\mathcal{N}_i^C(t)$.

\paragraph{Interaction network and interaction probability matrix} The \emph{interaction network} represents all short-term interactions established between agents at each time step $t$. It is the basis for updating both the agent characteristics and the contact network at each time step. We describe it as an undirected and simple time-dependent graph $G^I(t)$. It can be represented by its adjacency matrix $\mathbf{A}^I(t)$. The neighbourhood of agent $i$ in the interaction network is denoted by $\mathcal{N}_i^I(t)$.

Each entry $\pi_{ij}(t)$ of the \emph{interaction probability matrix} $\Pi(t)$ gives the probability that two agents $i,j$ will interact in time step $t$. For computing $\pi_{ij}(t)$, the social distance between two agents $i,j$ is measured by the shortest path length $d_{ij}^C(t-1)$ between them in the contact network at time $t-1$ (the minimum number of steps needed to reach agent $i$ from agent $j$ over the contact network). Empirical results reveal that social influence decays approximately exponentially with $d_{ij}^C$~\cite{christakis_fowler12}. Following these findings reporting a so-called ``three degrees of separation law'' of social influence, we define the interaction probability matrix as
\begin{align}
\pi_{ij}(t) = (\beta - \epsilon) \, \exp\left(-\frac{d_{ij}^C(t-1)-1}{\delta}\right) \widehat{L}(d_{ij})+ \epsilon, \label{eq:interaction_probabilities}
\end{align}
where $\epsilon$ is a baseline probability of interaction irrespective of the contact network. To account for the distribution of shortest path lengths between nodes, a normalisation factor $\widehat{L}(d_{ij})=L(1)/L(d_{ij})$ is introduced with $L(1)$ and $L(d_{ij})$ being the absolute number of shortest paths between nodes of length equal to 1 or $d_{ij}$, respectively. $\pi_{ij}(t)$ is scaled such that the probability of interaction between direct neighbours is always equal to the \emph{interaction probability scaling factor} $\beta$. Here we set $\beta = 0.8$ and $\epsilon=0.03$. The parameter $\delta$ gives the typical social distance for the exponential decay of interaction probability and is chosen as $\delta=2$ based in line with empirical evidence~\cite{christakis_fowler12}.

\paragraph{Proximity matrix} The \emph{proximity matrix} $\mathbf{P}(t)$ with elements $P_{ij}(t)$ measures the social proximity of two agents $i,j$. If social proximity is large, both agents are more likely to establish or maintain a contact. Hence, the proximity matrix is used in updating the contact network (see below). In our case of a single binary individual characteristic, the social proximity $P_{ij}(t)$ of two agents is only determined by the smoking behaviour $s_{i}(t)$ and $s_{j}(t)$ and their initially prescribed background proximity $B_{ij}$ describing rigid social ties such as family relationships and other factors that are not explicitly included in the model. We choose a simple linear relationship
\begin{align}
P_{ij}(t) = \alpha \left(1 - |s_i(t) - s_j(t)|\right) + (1 - \alpha) B_{ij}, \label{eq:proximity_matrix}
\end{align}
where $\alpha$ is a weight parameter balancing the influence of smoking behaviour and background proximity. Here, we choose $\alpha=$0.2 to allow for a typical network mobility of between one and two degrees in the Watts-Strogatz network that underlies the background proximity generation.

The \emph{background proximity matrix} $\mathbf{B}$ with elements $B_{ij}$ is constructed on the basis of a Watts-Strogatz small-world network~\cite{Watts1998a} with $N$ nodes, mean degree $z=10$ and a rewiring probability $p_w=0.03$. The individual proximities $B_{ij}$ are derived as a linear combination of the social distance $d_{ij}^\text{WS}$ in a realisation of a Watt-Strogatz random network and a uniformly distributed stochastic component $\zeta \in [0,1)$. This choice allows to emulate the typical small-world property of empirical social networks. $B_{ij}$ is derived as 
\begin{align}
B_{ij}= 1 - 0.1 (d_{ij}^{WS} - 1) - 0.1 \zeta.
\end{align}
After computation of $B_{ij}$ using the above formula, all entries with $B_{ij}<0.2$ are reset to a minimum value of $0.2$, which is in line with the assumption that for very high degrees of separation, no further meaningful distinction can be motivated.

\paragraph{Model dynamics}

The temporal update scheme describes the dynamics of the main variables of interest in the model (Fig.~\ref{fig:scheme}): smoking behaviour $s_i(t)$ and contact network $G^C(t)$. The model evolves in discrete time steps. It is deterministic with the exception of the stochastic generation of the interaction network from the interaction probability matrix and the stochastic switching of the agents' smoking behaviours in each time step. After initialisation, the algorithm proceeds from step 1 to step 6 in the full co-evolutionary setup and then starts again at step 1. Modified model dynamics implemented to isolate the effects of specific mechanisms in the model are described below.

\paragraph{Step 1: Calculate interaction probabilities based on social distance}

The interaction probability matrix $\mathbf{\pi}(t)$ is computed based on the social distance $d_{ij}^C(t-1)$ between agents $i,j$ in the contact network $G^C(t-1)$ from the previous time step $t-1$ following Eq.~\ref{eq:interaction_probabilities}.

\paragraph{Step 2: Generate interaction network}

The interaction network's adjacency matrix is randomly generated for all $i,j \in V$ independently. An interaction takes place with probability $\pi_{ij}(t)$ corresponding to setting $A_{ij}^I(t) = 1$, while no interaction takes place with probability $1-\pi_{ij}(t)$ leading to $A_{ij}^I(t) = 0$.

\paragraph{Step 3: Change agent characteristics (social influence / induction step)}

In the considered case of a single binary individual characteristic (smoking behaviour), social influence reduces to a probabilistic switching of smoking behaviour similar to the flipping of spins in an Ising model in physics~\cite{Castellano2009}. At time step $t$, the probability $p_i(t)$ to switch the smoking behaviour is assumed to depend to both the smoking disposition $\gamma_i(t-1)$ of agent $i$ and the average smoking behaviour in the agent's neighbourhood in the interaction network $G^I(t)$ at time $t$. For all agents $i$, the smoking behaviour is determined as follows. For a non-zero number of interactions $k_i^I(t) = \sum_{j=1}^N A_{ij}^I(t)$:
\begin{itemize}
\item If $s_i(t-1) = 0$ (non-smoker):
\begin{align}\label{eq:ising_nonsmoker}
p_i(t) = C \gamma_i(t-1) \sum_{j \in \mathcal{N}_i^I(t)} \frac{s_j(t-1)}{k_i^I(t)}
\end{align}
The smoking behaviour switches to ``smoker'' with probability $p_i(t)$, i.e. $s_i(t)=1$, and remains the same with probability $1-p_i(t)$, i.e. $s_i(t)=0$.

\item If $s_i(t-1) = 1$ (smoker):
\begin{align}\label{eq:ising_smoker}
p_i(t) = C (1-\gamma_i(t-1)) \sum_{j \in \mathcal{N}_i^I(t)} \frac{1-s_j(t-1)}{k_i^I(t)}
\end{align}
The smoking behaviour switches to ``non-smoker'' with probability $p_i(t)$, i.e. $s_i(t)=0$ and remains the same with probability $1-p_i(t)$, i.e. $s_i(t)=1$.
\end{itemize}
If no interactions take place for agent $i$ ($k_i^I(t)=0$), we set $s_i(t)=s_i(t-1)$.

The \emph{smoking behaviour switching probability scaling factor} $C$ scales the switching probability $p_i(t)$ of the smoking behaviour. $C$ controls the amplitude of equilibrium stochastic noise of the smoking behaviour that is introduced by the Ising-like implementation. Here we set $C=0.1$.

\paragraph{Step 4: Calculate proximity matrix}

The proximity matrix $\mathbf{P}(t)$ is computed from the current agent characteristics (smoking behaviour $s_i(t)$) and background proximity matrix $\mathbf{B}$ according to the diagnostic relationship given in Eq.~\ref{eq:proximity_matrix}.

\paragraph{Step 5: Update contact network}

New ties in the contact network $G^C(t)$ can only be established between agents that interacted in the same time step. In contrast, potentially any edge may disappear from the contact network depending on the outcome of the following update scheme. Let $U_i(t)$ be the list of neighbours $\mathcal{N}_i^I(t) \cup \mathcal{N}_i^C(t-1)$ of $i$ included in its interaction and contact neighbourhoods that is sorted in descending order of social proximity $P_{ij}(t)$.
The number of potential contacts of agent $i$ is determined by the agent's degree preference $q_i$. Specifically, the agent's potential contact neighbourhood at time $t$ consists of the set $T_i(t)$ of the first $q_i$ entries of $U_i(t)$.
Additionally, we require bidirectionality of contacts. This implies that only those agents can establish or maintain an contact relationship at time $t$ that are included in each other's potential contact lists $T_i(t)$. Thus, the contact network at time $t$ is derived as follows:
\begin{align}
A_{ij}^C(t) = \left\{ \begin{array}{rl}
1 & \mbox{if\,\,\,} i \in T_j(t) \wedge j \in T_i(t), \\ 
0 & \mbox{otherwise.}  \\
\end{array}\right.
\end{align}
This means that a new contact relation can only be formed if the corresponding social proximity value is large enough to enter the potential contact neighbourhood of both involved agents. On the contrary, a contact is lost if either alter is not element of ego's own list of potential contacts or if ego is not element of alter's own list, or if both is the case. Thus ego can loose a contact actively (by dropping alter) or passively (by being dropped by alter).

Importantly, we do not derive ``second best'' solutions by iteratively updating the potential contacts after the bidirectionality check. We account for this refinement implicitly via the iterative network update dynamics cycle in the model (Fig.~\ref{fig:scheme}).

\paragraph{Step 6: Apply external forcing (optional)}

External forcing can change agent properties and other system parameters. In our smoking case study, we change the smoking dispositions $\gamma_i(t)$ of agents over time according to prescribed initial and target distributions to emulate the effects of changing values, political and health campaigns, etc. For all time steps, we ensure that the set $\{\gamma_i(t)\}_i$ is consistent with being drawn from a parabolic probability distribution of the form $y(x; t)= a(t) (b(t) - x)^2 + c(t)$ for $x \in [0,1]$ with parameters implicitly defined by the conditions $\int_{0}^{1}y(x;t) dx=1$, $y(x=0;t)=C_1$ and $y(x=1;t)=C_2(t)$. Specifically, the initial $\{\gamma_i(0)\}_i$ are drawn from a bimodal and symmetric distribution $y(x;t=0)$ with $C_2(t)=C_1$. 

Then, the target distribution is changed over time by gradually reducing the parameter $C_2(t)$. To compute the set of smoking distributions $\{\gamma_i(t)\}_i$ at time step $t$, the previous set $\{\gamma_i(t-1)\}_i$ is stochastically transformed by stepwise addition of two-tailed log-normal distributed noise $\epsilon$ that is linearly weighted by  the deviation from the target distribution. Noise is added iteratively until a Kolmogorov-Smirnoff criterion with significance level 90 \% of $\{\gamma_i(t)\}_i$ being drawn from the target distribution $y(x; t)$ is fulfilled. By this procedure, individual $\gamma_i$ are modified following a Markov process, whereas the overall system property, in this case the smoking disposition distribution, is externally controlled. Using this procedure, the randomly sampled initial set of smoking dispositions $\{\gamma_i(0)\}_i$ following a bimodal distribution ($C_2(t=0)=C_1$) is gradually transformed into a sample $\{\gamma_i(t_f)\}_i$ following a quasi unimodal distribution ($C_2(t=t_f)=C_{f}$; see Fig.~\ref{fig:scheme}).

\paragraph{Modelling protocol} Model runs proceed in three steps: 
(i)~The interaction network is initialised as $A_{ij}^I(0)=1$ for all pairs $i,j$. In the following, the  initial contact network $A_{ij}^C(0)$) is established based on the fully connected interaction network. Smoking behaviour $s_i(0)$ is initialised consistently with the initial smoking attitude $\gamma_i(0)$ as
\begin{align}
s_i(0) = \Theta\left(\gamma_i(0) - \frac{1}{2}\right),
\end{align}
where $\Theta(\cdot)$ is the Heaviside function. 
(ii)~The system is then integrated without applying external forcing for 200 time steps to a quasi-equilibrium state. We choose system parameters interaction probability scaling factor $\beta=0.8$ and smoking behaviour switching probability scaling factor $C=0.1$ to limit the system's internal noise level in equilibrium. More specifically, this choice guarantees that the maximum deviation from the median number of smokers in equilibrium that is induced by the stochastic dynamics of the model is smaller than 5\,\% of the population size $N$.
(iii)~The system is then further integrated under continuous application of the stochastic external policy forcing acting on the smoking dispositions $\gamma_i(t)$.

\paragraph{Partial models} Besides the fully \emph{coupled} model described above, we study three additional partial models that focus on a subset of processes of behaviour formation (Tab.~\ref{table}): (i) an \emph{interaction} model focussing on local social influence by assuming a static contact network (omitting step 5), (ii) a \emph{network} model not considering local social influence, but inducing behavioural change only whenever the exogenously modified smoking disposition $\gamma_i(t)$ of an individual $i$ crosses a threshold of 0.5 (modifying step 3), (iii) a \emph{mean-field} model, where the agents react to the mean-field effect of the average smoking prevalence $S(t) / N$ instead of their local neighbourhood in the social influence process (modifying step 3).

\subsection*{Network metrics}

\paragraph{Eigenvector centrality} The eigenvector centrality $c_i(t)$ of agent $i$ (also referred to simply as \emph{centrality} above) is a non-local centrality measure implicitly defined to be proportional to the sum of $i$'s contact neighbours' eigenvector centralities~\cite{Bonacich1972}. An agent has a high centrality if it is connected to many high centrality neighbours in the contact network that also have many high centrality neighbours. This implies that large values of eigenvector centrality $c_i(t)$ are observed in high density cliques or substructures embedded within the contact network. $c_i(t)$ is given by the $i$-th component of the leading eigenvector (associated to the largest eigenvalue) of the contact network's adjacency matrix $\mathbf{A}^C(t)$ at time $t$ and is computed by applying the \texttt{evcent} method from the \texttt{igraph} package~\cite{Csardi2006}.

\paragraph{Conditional probability of smoking} The conditional probability that a randomly drawn agent $i$ (ego) smokes given that another agent $j$ (alter) randomly drawn from a neighbourhood shell at social distance $d_{ij}^C(t) = d$ smokes is defined as
\begin{align}
P(d; t) = \frac{\frac{1}{S(t)}\sum_{i \in \mathcal{S}(t)} \frac{1}{N_{i,d}(t)}\sum_{j \in \mathcal{N}_{i,d}(t)} s_j(t)}{S(t) / N}.
\end{align}
Here, $\mathcal{S}(t)$ denotes the set of smokers at a given time-step and $S(t) = |\mathcal{S}(t)|$ is the total number of smokers. Similarly, $\mathcal{N}_{i,d}(t)$ is the set of agents at a social distance $d$ (measured by distance on shortest paths) from agent $i$ in the contact network and $N_{i,d}(t) = |\mathcal{N}_{i,d}(t)|$ is the total number of agents in this set. In our study, $P(d; t)$ is employed as a measure of the mean effect of social distance in the contact network on smoking behaviour~\cite{Christakis2008}.



\section*{Acknowledgements}


The reported research was conducted within the scope of the COPAN flagship project on co-evolutionary pathways at the Potsdam Institute for Climate Impact Research (http://www.pik-potsdam.de/copan). The authors acknowledge support by the German National Academic Foundation. D.E. acknowledges support by an Amazon in Education grant. J.F.D. thanks the Stordalen Foundation (via the Planetary Boundary Research Network PB.net) and the Earth League's EarthDoc program for financial support. The authors gratefully acknowledge the European Regional Development Fund (ERDF), the German Federal Ministry of Education and Research and the Land Brandenburg for supporting this project by providing resources on the high performance computer system at the Potsdam Institute for Climate Impact Research. Danilo Bzdok, Guillaume Dumas, Jobst Heitzig, Reik V. Donner and Wolfram Barfuss are acknowledged for valuable comments and discussions.

\section*{Author contributions statement}

C.F.S., J.F.D., D.E. and A.L. designed the research. C.F.S., J.F.D. and D.E. performed the research and wrote the manuscript. All authors reviewed the manuscript. 

\section*{Additional information}



The authors declare no competing financial interests.

\end{document}